\documentclass[12pt]{article}
\usepackage{graphicx}
\begin{document}
\rightline{NKU-2013-SF3}
\bigskip

\newcommand{\be}{\begin{equation}}
\newcommand{\ee}{\end{equation}}
\newcommand{\noi}{\noindent}
\newcommand{\refb}[1]{(\ref{#1})}
\newcommand{\ra}{\rightarrow}
\newcommand{\bi}{\bibitem}

\begin{center}
{\Large\bf String black hole: Can it be  a  particle accelerator ?}

\end{center}
\hspace{0.4cm}
\begin{center}
Sharmanthie Fernando \footnote{fernando@nku.edu}\\
{\small\it Department of Physics \& Geology}\\
{\small\it Northern Kentucky University}\\
{\small\it Highland Heights}\\
{\small\it Kentucky 41099}\\
{\small\it U.S.A.}\\

\end{center}

\begin{center}
{\bf Abstract}
\end{center}

In this paper we have studied the possibility of the center-of-mass energy of two particles colliding near the horizon of a static charged black hole in string theory. Various cases corresponding to the electric charge and the angular momentum of the particles were considered. The studies were done for the general black hole as well as for the extreme black hole. There were two scenarios where the center-of-mass energy  reach very large values if the appropriate properties of the particles are chosen.

\hspace{0.7cm}

{\it Key words}: static, charged,  string, accelerator, black hole

%%%%%%%%%%%%%%

\section{ Introduction}

Ban$\tilde{a}$dos, Silk and West showed that when two particles collides near the horizon of a Kerr black hole, that the center of mass energy could become arbitrarily high \cite{bana}. Hence  black holes could act like particle accelerators and be a platform to probe Plank scale physics. This idea, called BSW effect (called after the three authors) was first studied for the extreme Kerr black hole. Then it  was extended by Grib et.al  to show that high energy can be obtained for non-extremal Kerr black holes when there are multiple collisions \cite{grib1}.  In another paper, Grib and Parlov discussed how the center of mass energy of colliding particles could grow  on the ergosphere of a rotating black hole \cite{grib2}. Harada and Kimura computed the near horizon collision of two particles which were falling from inner most stable circular orbit (ISCO) of a Kerr black hole \cite{harada}. Extension of BSW effect to many other black holes exist in the literature. Li et.al showed that the Kerr-(anti)-de Sitter black hole also could act as a particle accelerator \cite{li}. Studies were done by Zhu et.al. on getting infinite center of mass energy of two charged particles in a general stationary charged black hole \cite{zhu}. Particle acceleration on the background of the Kerr-Taub-NUT black holes were studies by Liu et.al. \cite{liu2}. Effects of ultra-high energy collisions of particles near the horizon of the BTZ black hole was studies by Yang et.al \cite{yang}. Rotating charged cylindrical black holes as particle accelerators were studied by Said and Adami \cite{said}.

All the papers we referred above were for considering particle acceleration around rotating black holes. Rotation was essential to achieve the unlimited energy for the center of the mass. However, in an interesting paper,  Zaslavskill \cite{zasl1} presented that similar effects can be achieved for non-rotating charged black holes as well. Studies of particle collision near the cosmological horizon around the Reissner-Nordstr$\ddot{o}$m-de Sitter black hole was done by Zhang and Gao \cite{gao}. It was shown that the center of mass energy will become infinite closer to the cosmological horizon. Partil and Joshi studied the center of mass energy of two particles colliding near a naked singularity by studying the Janis-Newmann-Winicour space-time and showed that the center of mass energy could become large near the naked singularity \cite{patil}.

In this paper, we study the BSW effect around a charged black hole in string theory. Given the fact that string theory is the leading candidate to unify General Theory of Relativity and quantum mechanics, the black hole solution of the low energy string effective action also take a fundamental importance in understanding the universe. The black hole we consider is the static charged 4 dimensional black hole in heterotic string theory. The action considered is given by,

\begin{equation} \label{action}
S = \frac{1}{ 16 \pi} \int d^4x \sqrt{-g} \left[ R - 2 (\bigtriangledown \Phi )^2 -
e^{-2  \Phi} F_{\mu \nu} F^{\mu \nu}  \right]
\end{equation}
Here, $ \Phi$ is the dilaton field, $R$ is the scalar curvature and $F_{\mu \nu}$ is the Maxwell's field strength. The static charged black hole solutions for the above action were found independently by two groups: Gibbons and Maeda \cite{gib} and Garfinkle, Horowitz and Strominger \cite{gar}. The black hole solution is well known as  the Gibbons-Maeda-Garfinkle-Horowitz-Strominger (GMGHS) black hole.

Some aspects of collisions of particles in this black hole were discussed by Mao et.al in \cite{mao}. In the current paper, we will do a thorough analysis of all possible scenarios. Particle collisions of another black hole in string theory, which is known as the Sen black hole were studied in the same context by Wei et.al \cite{wei}. The Sen black hole has spin and also has other fields in the back ground. Such fields were not included in the action $\refb{action}$. The paper is organized as follows: In section 2, an introduction is given for the GMGHS black hole. The geodesics and the center-of-mass energy is  presented in section 3. In section 4, the general black hole is studied for particle collisions. In section 5, the extreme black hole is studied. In section 6, the conclusions are given.

%%%%%%%%%%%%%%%%%%%%%%%%%

\section{ Introduction to the GMGHS black hole}

The static charged black hole (GMGHS black hole) solutions to the action given in eq. $\refb{action}$ is given by,
\begin{equation} \label{black}
ds^2 = -f(r) dt^2 + f(r)^{-1} dr^2 +  R(r)^2 ( d \theta^2 + sin^2(\theta) d \phi^2)
\end{equation}
Here,
\begin{equation}
f(r) = \left(1-\frac{2M}{r}\right)
\end{equation}
\begin{equation}
R(r)^2 = r\left(r-\frac{Q^2} {M} \right) = r ( r -b)
\end{equation}
where, 
\begin{equation}
 b = \frac{Q^2}{M}
\end{equation}
The electric field strength and the dilaton field are given by,
\begin{equation}
{ F_{rt}=\frac{Q}{r^2};\qquad e^{2\Phi}
= 1-\frac{Q^2}{Mr}}
\end{equation}
There is an event horizon at $ r = 2M$ and $ r = Q^2/M$ surface is singular. The singularity is evident when computed the scalar curvature to be,
\be
\mathcal{R} = \frac{ b^2 ( 2 - 2 M)}{ 2 r^3 ( r - b)^2}
\ee
When  $Q^2 \leq 2M^2$, the singular surface is  inside the event horizon. When $ Q^2=2 M^2$, the singular surface  coincide with the horizon which is  the extremal limit where a transition between the black hole and the naked singularity occurs. For $ Q^2 > 2 M^2 $, the solution in eq.$\refb{black}$ becomes a naked singularity.

In comparison with the Schwarzschild black hole, both have event horizons at $ r = 2 M$. However, the GMGHS solution could exhibit a naked singularity for the appropriate values of $ Q$ and $M$. The temperature of both black holes are identical. It is given by,
\be
T = \frac{ 1}{ 8 \pi M}
\ee

%%%%%%%%%%%%%%%
\section{ The geodesics of the test particles and the center-of-mass energy around the GMGHS black hole}

To study the center-of-mass (CM) energy of two particles, first the geodesics and the four-velocity of the particles has to be derived.

\subsection{ Geodesics}

To derive the geodesics, we will follow the well known book by Chandrasekhar\cite{chandra}.  Here we will derive the equations of the motion of a charged test particle with electric charge per unit mass $e$. Such equations can be derived from the   Lagrangian equation,
\begin{equation} \label{lag}
{\cal{L}} =  - \frac{1}{2} \left( - f(r) \left( \frac{dt}{d\tau} \right)^2 +  \frac{1}{f(r)}\left( \frac{dr}{d \tau} \right)^2 + R(r)^2 \left(\frac{d \theta}{d \tau} \right)^2 + R(r)^2 sin^2 \theta \left( \frac{d \phi }{d \tau} \right)^2 \right) + 2 e A_t \frac{ dt}{ d \tau}
\end{equation}
Here, $\tau$ is the proper time for time-like geodesics ( or massive particles) and $ A_t = \frac{Q}{r}$  is the electric potential. The canonical momenta corresponding to each coordinate is given as,
\begin{equation} \label{pt}
p_t = \frac{d {\cal{L}} }{d \dot{t}} = f \dot{t} + \frac{ e Q}{r}
\end{equation}
\begin{equation}
p_{r} = - \frac{d {\cal{L}} }{d \dot{r}} = \frac{ \dot{r}}{f}
\end{equation}
\begin{equation}
p_{\theta}  = - \frac{d {\cal{L}} }{d \dot{\theta}} =  R(r)^2 \dot{\theta}
\end{equation}
\begin{equation} \label{pphi}
p_{\phi}  = - \frac{d {\cal{L}} }{d \dot{\phi}} =  R(r)^2 sin^2 \theta \dot{\phi}
\end{equation}

The GMGHS black hole  have two Killing vectors $\partial_t$ and $ \partial_{\phi}$. Hence, there are two conserved quantities along the motion of the particle which can be labeled as $E$ and $L$.  From eq.$\refb{pt}$ and eq.$\refb{pphi}$,  $E$ and $L$ are related to $f(r)$ and $R(r)$ as,
 \begin{equation} \label{pt2}
p_t = \frac{d {\cal{L}} }{d \dot{t}} = f \dot{t} + \frac{ e Q}{r} =  E
\end{equation}
\begin{equation} \label{pphi2}
p_{\phi}  = - \frac{d {\cal{L}} }{d \dot{\phi}} =  R(r)^2 sin^2 \theta \dot{\phi} = L
\end{equation}
We will consider the motion on the equatorial plane. Hence $\theta = \pi/2$, $\dot{\theta} =0$, \& $\ddot{\theta} =0$. From eq.$\refb{pt2}$ and eq.$\refb{pphi2}$,
 \begin{equation} \label{pt3}
\dot{t} = \frac{E -  \frac{ e Q}{r}}{ f(r)}
\end{equation}
\begin{equation} 
 \dot{\phi} = \frac{L}{R(r)^2}
\end{equation}
The fours velocity of the particles are given by $u^{\mu} = \frac{ dx^{\mu}}{d \tau}$.  We have already obtained $u^{t}, u^{\theta}$ and $ u^{\phi}$ in the above derivations. To find $u^{r} = \dot{r}$,  the normalization condition for time-like particles, $u^{\mu} u_{\mu} = -1$ can be used as,
\be \label{nor}
g_{tt} (u^t)^2 + g_{rr}(u^r)^2 + g_{\theta \theta} (u^{\theta})^2 + g_{\phi \phi} (u^{\phi})^2 = -1
\ee
By substituting $u^t, u^{\phi}$ and $u^{\theta}$ to eq.$\refb{nor}$,  one can obtain $u^r$ as,
\begin{equation} \label{rdot}
(u^r)^2=  \dot{r}^2   = -  V_{eff} 
\end{equation}
where $V_{eff}$ is the effective potential for the motion, given by,
\begin{equation} \label{pot1}
V_{eff} =   f(r) \left( 1 +  \frac{L^2}{R(r)^2} \right)   - \left( E - \frac{ e Q}{r}\right)^2
\end{equation}
If we impose the condition that $\dot{r} =0$ when $ r \ra \infty$, then $E = 1$ from eq.$\refb{pot1}$. Also, from eq.$\refb{pt3}$, we will assume 
\be \label{timeforward}
( E - \frac{ e Q}{r_h})  = ( 1 - \frac{ e Q}{r} ) >  0 
\ee
for all $ r > r_h$ so that the motion is forward in time out side the horizon $r_h$. Now, the four-velocity values can be written as,
\be
u^t = \dot{t} = \frac{(1 -  \frac{ e Q)}{r}}{ f(r)}
\ee
\be
u^{\phi} =  \dot{\phi} = \frac{L}{R(r)^2}
\ee
\be
u^{\theta} = \dot{\theta} =0
\ee
\be
u^r = \dot{r} = \sqrt{ - V_{eff} } =  \sqrt{ f(r) \left( 1 +  \frac{L^2}{R(r)^2} \right)   - \left( 1 - \frac{ e Q}{r}\right)^2 }
\ee

\subsection{ Two particles collisions: Center-of-Mass energy ($E_{CM}$)}

In this section we will present the CM energy of two particles with four-velocity $u_1^{\mu}$ and $u_2^{\mu}$. We will assume that both have rest mass $m_0=1$. The CM energy is given by,
\be
\hat{E}_{cm} =  2 m_0^2 ( 1 - g_{\mu \nu} u_1^{\mu} u_2^{\nu} )
\ee
in the rest of the paper we will compute $\hat{E}_{cm}/2 = E_{cm}$ which is given by,
$$
E_{cm} =  \left( 1 + \frac{ ( 1 - \frac{ e_1 Q}{r}) ( 1 - \frac{ e_2 Q}{r})}{ f(r)} -  \frac{ L_1 L_2} { R(r)^2 } \right) + 
$$
\be \label{cm}
\sqrt{ \frac{ ( 1 - \frac{ e_1 Q}{r})^2}{ f(r)} - \left( 1 + \frac{L_1^2}{R(r)^2}\right) }  \sqrt{ \frac{ ( 1 - \frac{e_2 Q}{r})^2}{ f(r)} - \left( 1 + \frac{L_2^2}{R(r)^2}\right) } 
 \ee
In the subsequent sections, various cases of collisions will be considered. The mass of the black hole is fixed at $1$ so that the horizon is at $r_h$ in the rest of the paper. Hence $f(r) = (1- \frac{2}{r})$.
%%%%%%%%%%%%%%%%%%%%%
\section{ CM energy for two particle collision around GMGHS black hole with $ b < 2 M$}

Here, we will analyze the two particle collisions for variety of cases.

%%%%%%%%%%%%%%%%%%%%%%%%%%

\subsection{ CM energy when $e_1 = 0, e_2  = 0, L_1 \neq 0, L_2 \neq 0$}

In this case, the effective potential becomes,
\be \label{potneutral}
V_{eff} = - \frac{2}{r} + \left( 1 - \frac{2}{r} \right)  \frac{ L^2}{R^2(r)}
\ee
 
 From eq.$\refb{potneutral}$, it is clear that $V_{eff} < 0$ for  $ r = r_h =2$. For $ r \ra \infty$, $V_{eff} \ra 0$. Depending on the values of $L$ the potential could be positive or negative for $ r > r_h$.  Since $ \dot{r}^2 + V_{eff} =0$, $V_{eff}$ has to be negative for all $r > r_h$ for a particle to fall into the black hole. Therefore, it is important to study the behavior  of the potential and see when it becomes positive. The potential is plotted for various values in the Fig.1. From the figure, it is clear that there exists a critical angular momentum that separates the behavior of the potential. When $ L > L_c$, there are turning points for the potential and the particle will not be able to reach the horizon. When $ L = L_c$, the potential is negative and is zero at a critical radius $r_c$. When $ L < L_c$, the $V_{eff}$ is negative for all $r > r_h$. Hence, for the collisions to occur,  $|L| < L_c$. When $ L > L_c$, there are two turing points for $V_{eff} =0$ at
  \be \label{root}
 r = \frac{ 1}{4} \left ( 2 b + L^2 \pm \sqrt{ ( 2 b + L^2)^2 - 16 L^2} \right)
 \ee
 When $ L = L_c$, $V_{eff} =0$ at $ r = r_c$ which is the degenerate root of the eq.$\refb{root}$ given by,
 \be
 r_c = 2 + \sqrt{ 4 - 2 b}
 \ee
 The corresponding $L_c$ is given by,
 \be
 L^2_c = ( 8 - 2 b ) + 4 \sqrt{ 4 - 2 b}
 \ee

  \begin{center}
\scalebox{.9}{\includegraphics{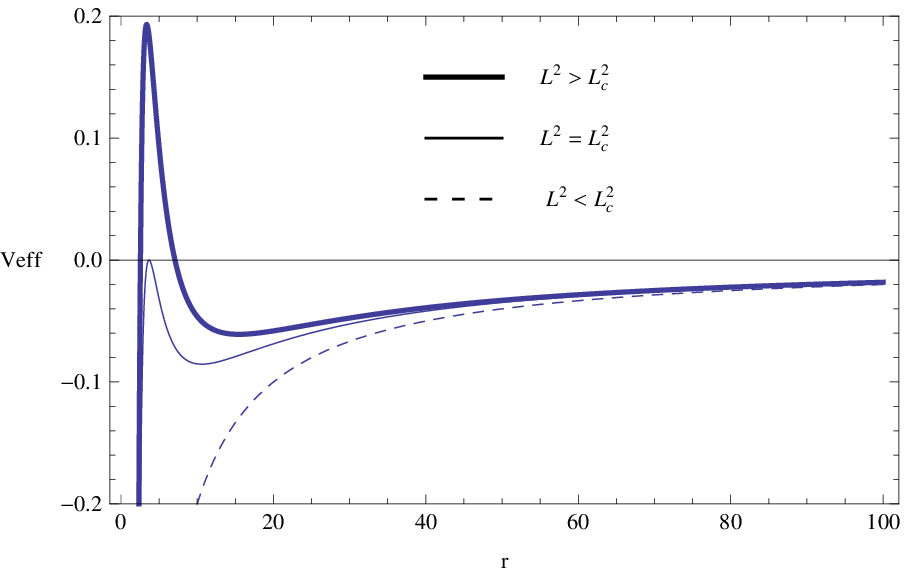}}\\
\vspace{0.1cm}
\end{center}
Figure 1. The graph shows the graph  of $V_{eff}$ vs $r$.  Here, $M =1, Q =0.707, L \neq 0$ and $ r_h = 2$.\\

 The CM energy for this case is given by,
 
$$
E_{cm} =  \left( 1 + \frac{  1 }{ f(r)} -  \frac{ L_1 L_2} { R(r)^2 } \right) + 
$$
\be
\sqrt{ \frac{ 1 }{ f(r)} - \left( 1 + \frac{L_1^2}{R(r)^2}\right) }  \sqrt{ \frac{  1}{ f(r)} - \left( 1 + \frac{L_2^2}{R(r)^2}\right) } 
 \ee
If one takes the limit, $r \ra 2$, $E_{cm}$ becomes,
\be
E_{cm} = \frac{ 1}{ 2 ( 2 -b)} \left( ( L_1 - L_2)^2 - 8 b + 16 \right)
\ee

One can note that there is no possibility for $E_{CM}$ to be infinite. Note that $ b <2$ since the singularity is hidden behind the horizon $r =2$. Also, when the angular momentum of the particles are zero, $E_{CM}$ becomes $2$ which is similar to the value obtained (upto a constant) for the Schwarzschild black hole in \cite{bana}. When there is no angular momentum, the effective potential is same as it is for the Schwarzschild black hole and there should not be any difference in the behavior of the particle.

When charge of the black hole goes to zero,  $b \ra 0$ and the black hole becomes the Schwarzschild black hole. Then,
\be
E_{cm} \ra \frac{ 1}{ 4} \left( ( L_1 - L_2)^2  + 16 \right)
\ee
Here $E_{CM}$ is finite as discussed in \cite{bana}. It will become maximum when $L_1$ and $L_2$ are opposite in sign.

%%%%%%%%%%%%%%%%%%%%%%%%%

\subsection{ CM energy when $e_1 \neq 0, e_2 \neq 0, L_1 = L_2 =0$} \label{cmforlzero}

In this case, the effective potential for a particle with charge becomes,
\begin{equation} \label{pot2}
V_{eff} =    \left(1 - \frac{2}{r}  \right)   - \left( 1 - \frac{ e Q}{r}\right)^2 = \frac{(eQ)^2}{r^2} + \frac{ 2(1  -  e Q)}{ r}
\end{equation}
From   eq.$\refb{pot2}$, it is clear that $V_{eff} <0$  at the horizon, $r_h = 2$ for any value of $ eQ$. However, depending on the value of $eQ$, the potential could be positive or negative for $r > r _h$. Since $ \dot{r}^2 + V_{eff} = 0$,  $ V_{eff}$ has to be negative  for all $r > r_h =2$ for a particle to fall to the black hole from a large value of $r$. In fact, for large $eQ$ values, $V_{eff} >0$ for some values of $r$. For small values of $eQ$, $V_{eff} <0$ for all $r$. This behavior is represented in the Fig.2. The question is what is the value of $eQ$ which separates the two regions. One can compute the root of $V_{eff} =0$ as,
\be
r = \frac{ ( eQ)^2}{ 2 ( - 1 + e Q)}
\ee
When $eQ < 1$, the root is negative leading to  $V_{eff} <0$ for all $r > r_h$.   The potential in this case behaves as the dotted curve in Fig.2. When $eQ >1$,  $V_{eff} $ has a positive root, which is shown by the thick curve in Fig. 2.  Hence for the particle to fall into the black hole, $e Q < 1$. 

\begin{center}
\scalebox{.9}{\includegraphics{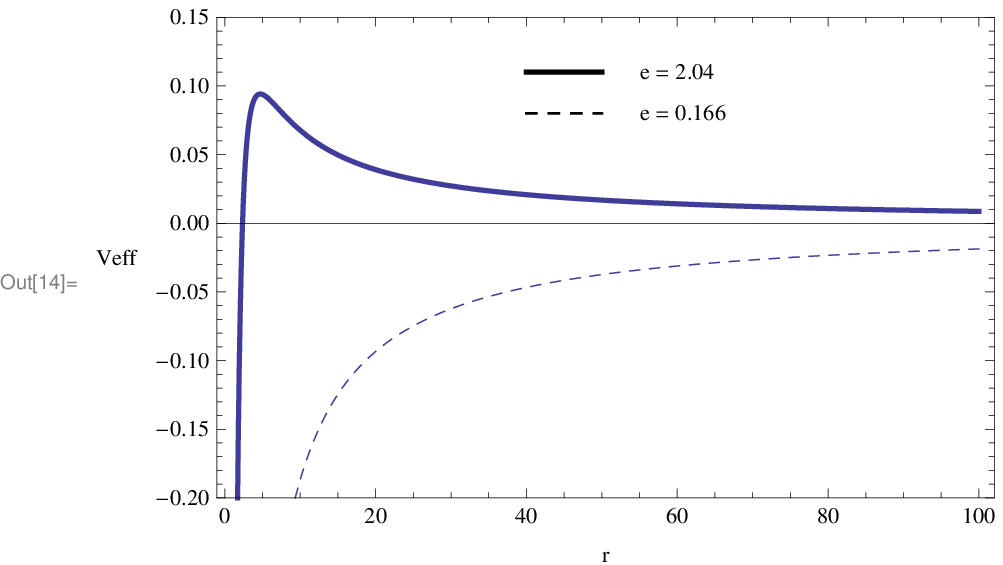}}\\
\vspace{0.1cm}
\end{center}
Figure 2. The graph shows the graph  of $V_{eff}$ vs $r$. The  electric charge of the particle is different for the two curves. Here, $M =1, Q =0.707, L =0$ and $ r_h = 2$.\\

Now, in this case, the CM energy $E_{CM}$ in eq.$\refb{cm}$ becomes,

$$
E_{cm} =  \left( 1 + \frac{ ( 1 - \frac{ e_1 Q}{r}) ( 1 - \frac{ e_2 Q}{r})}{ f(r)} \right) + 
$$
\be 
\sqrt{ \frac{ ( 1 - \frac{ e_1 Q}{r})^2}{ f(r)} -  1 }  \sqrt{ \frac{ ( 1 - \frac{e_2 Q}{r})^2}{ f(r)} -  1 } 
 \ee
Here $e_1$ and $e_2$ are the charges of each particle. Note that $( 1 - \frac{e_1 Q}{r}) >0$ and $( 1 - \frac{e_2 Q}{r}) >0$ for all $r > r_h$ from the condition in eq.$\refb{timeforward}$.When $r \ra r_h$, $f(r) \ra 0$. When $E_{CM}$ is expanded around $f(r) =0$, the expression becomes,
\be
E_{CM} = 1 + \frac{ 1}{2} \left[ \left( \frac{ 2 - e_1 Q)}{ 2 - e_2 Q} \right) + \left( \frac{ 2 - e_2 Q)}{ 2 - e_1 Q} \right) \right]
\ee
From the above expression, the only way $E_{CM}$ could go to infinity would be either $e_1 Q = 2$ or $e_2 Q =2$. But from the discussion on the effective potential,  $e_1Q < 1$ and $e_2 Q <1$ for either particles to reach the horizon. Hence, $E_{CM}$ will not reach  infinity in this case. In the paper by Mao etlal. \cite{mao}, the $E_{CM}$ was computed for this particular case and came to the conclusion that $ E_{CM}$ will become large. However, potential was not studied in that case to see that the particle will not reach the horizon for this to occur.

%%%%%%%%%%%%%%%%%%
\subsection{ CM when $e_1 \neq 0, e_2 \neq 0, L_1 \ne 0, L_2 \neq 0$ }

For this case, the potential is given by eq$\refb{pot1}$. The potential is plotted in the Fig.(3) for various values of $ L$. 
\begin{center}
\scalebox{.9}{\includegraphics{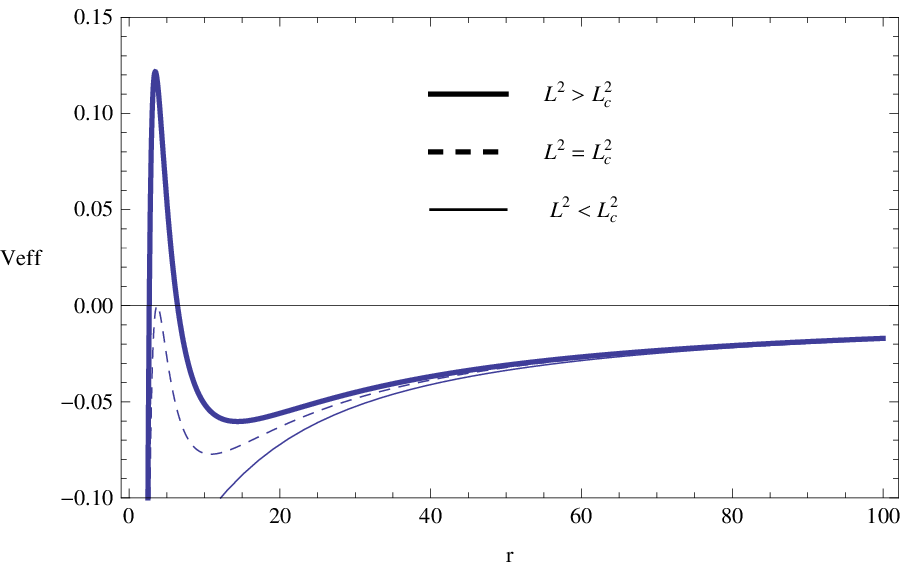}}\\
\vspace{0.1cm}
\end{center}
Figure 3. The graph shows the graph  of $V_{eff}$ vs $r$.  Here, $M =1, Q =0.707, e=0.1$ and $ r_h = 2$. For these values, $L_c = 3.598$ and $r_c = 3.733$.\\

From the graph, when the angular momentum is larger than a critical value (given by the thick graph), the potential is positive in certain regions. Hence, the potential has turning points. When the angular momentum is smaller than a critical value (given by the light curve), the potential is always negative. When the angular momentum is equal to the critical value, the potential is negative and zero at a critical radius $r_c$ (given by the dashed curve). As we have explained before, for the particle to fall into the black hole the potential should be negative. Hence, the angular momentum has to be smaller or equal to the critical value $L_c$. The way to find the critical values $ L_c$ and $r_c$ is to impose the condition,
\be \label{critical}
V_{eff}(r)=0 \hspace{1 cm} \ V'_{eff}(r) =0
\ee
It is not possible to give an analytical expressions for $r_c$ and $L_c$. However, we have found them numerically for various values of $M, b, e$ given in Fig.4 and Fig.5.

\begin{center}
\scalebox{.9}{\includegraphics{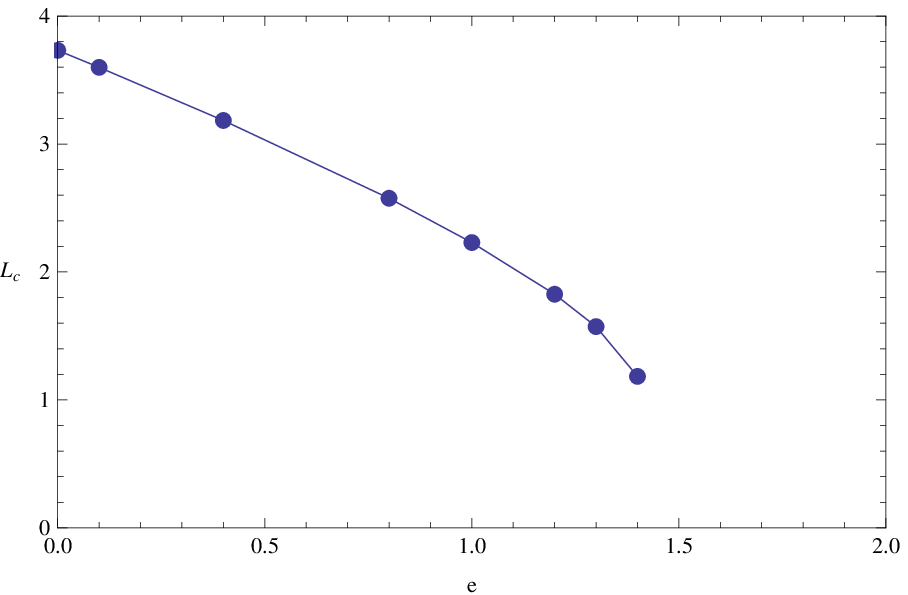}}\\
\vspace{0.1cm}
\end{center}
Figure 4. The graph shows the graph  of $L_c$ vs $e$. Here, $M =1, Q =0.707$ and $ r_h = 2$.

\begin{center}
\scalebox{.9}{\includegraphics{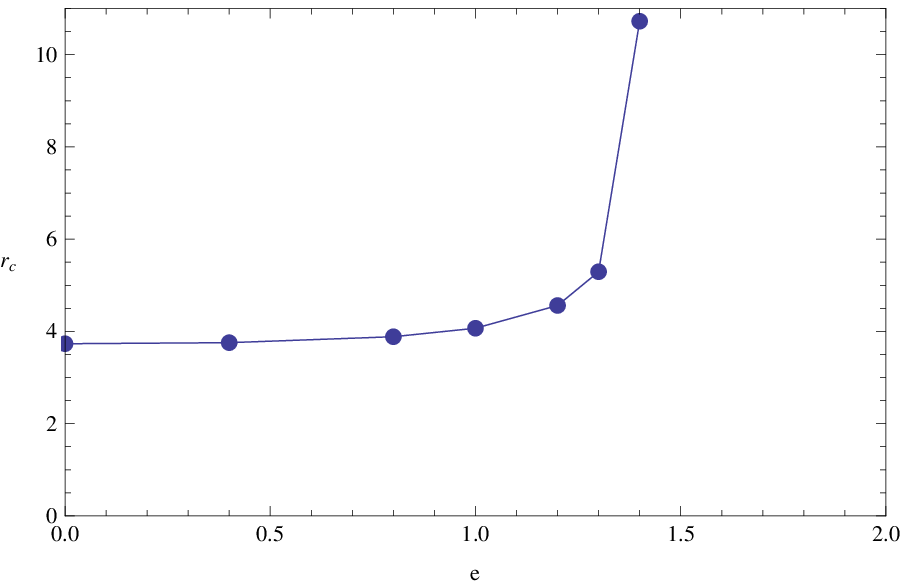}}\\
\vspace{0.1cm}
\end{center}
Figure 5. The graph shows the graph  of $r_c$ vs $e$.  Here, $M =1, Q =0.707$ and $ r_h = 2$. \\

Now  we can study the $E_{CM}$ closer to the horizon for this case. When the expression in eq.$\refb{cm}$ is expanded around $ f(r) \ra 0$, it simplifies to
\be
E_{CM}= \frac{ 1} { 2}  \left[  \left(  \frac{ L_1^2}{ n} + 1 \right) \left( \frac{ 2 - e_1 Q}{ 2 - e_2 Q} \right) +  \left(\frac{L_1^2}{ n} + 1 \right) \left( \frac{ 2 - e_2 Q}{ 2 - e_1 Q} \right) + 2 - \frac{ 2 L_1 L_2}{n}  \right]
\ee
Here, $n =  R^2(2) = 2 ( 2 - b)$ which is finite since $ b < 2$.  By observing the above expression, one can conclude that the only possibility for it to be infinite is for either $ e_1 Q =2 $ or $ e_2 Q = 2$. When we solved for $L_c$ and $r_c$ numerically,  what we observed was that $ L_c$  and $r_c$ becomes complex  when $eQ$ approach 1. By  changing the values of $b$ for other values, similar behavior was observed. Hence, when $eQ \ra 2$, both $ r_c$ and $L_c$ becomes complex numbers. Therefore, $eQ \ra 2$ is not a possible limit for the particle to fall into the black hole and $E_{CM}$ will not approach infinity in this case.

%%%%%%%%%%%%%%%%%%%%%%%%
\section{ Extreme black hole}

Extreme black hole occurs when $ 2 M = b$ in the metric in eq.$\refb{black}$. Hence the horizon and the singularity coincide at $ r = 2$ leading to,
\be
f(r ) = 1 - \frac{ 2}{ r} ; \hspace{1 cm}  R^2(r) =  r ( r -2)=  r^2 f(r)
\ee
There are discussions about the extreme GMGHS black hole in \cite{pra} and \cite{horo}. We will analyze several different scenarios when the particle collision could occur.

\subsection{ $e = 0$ and $L =0$}

In this case the effective potential becomes,
\be
V_{eff} = - \frac{ 2 M}{r}
\ee
It is clear that a particle will definitely fall towards the singularity. The CM energy in this case is given by,
\be
E_{CM} = 1 + \frac{1}{ f(r) } - \sqrt{ - 1 + \frac{ 1}{ f(r) }} \sqrt{ - 1 + \frac{ 1}{ f(r) }} 
\ee
The above expression simplifies when expanded around $f(r) \ra 0$ to be 2. Hence $ E_{CM}$ is finite similar to the collision around the horizon at the  Schwarzschild black hole \cite{bana}.

%%%%%%%%%%%%%%%

\subsection{ $e =0$ and $ L \neq0 $}

In this case, the effective potential becomes,
\be
V_{eff} = - \frac{ 2}{r} +\frac{ L^2}{r^2}
\ee
$V_{eff}$ will have a turning point at $ r = \frac{L^2}{2}$ as given in Fig.6.

\begin{center}
\scalebox{.9}{\includegraphics{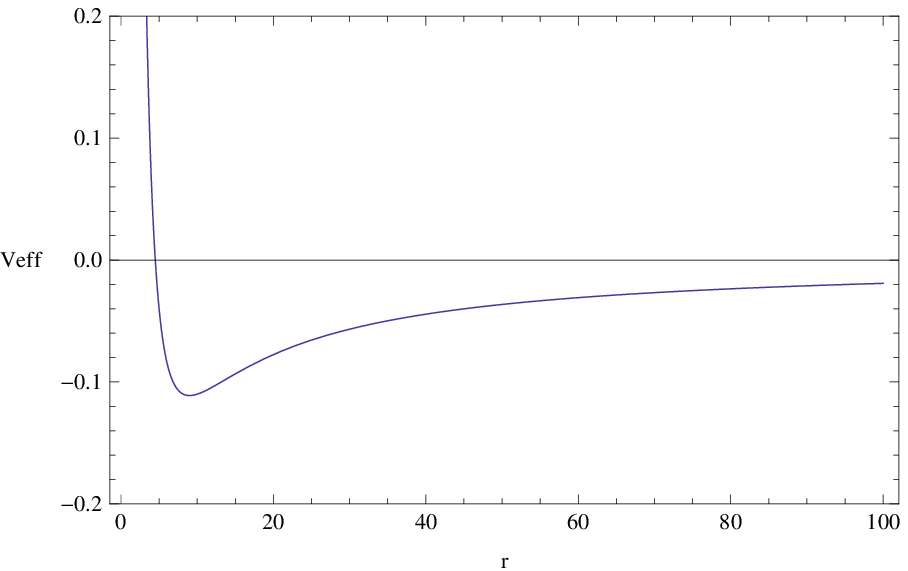}}\\
\vspace{0.1cm}
\end{center}
Figure 6. The graph shows the graph  of $V_{eff}$ vs $r$ for the extreme black hole with $M = 1$ and $ b = 2$ with angular momentum.

Since the horizon is at $ r =2$, if $ L^2 > 4$, the particle will not reach the horizon. If $ L^2 < 4$ then the particle can reach the horizon. The CM energy for this case is,
$$
E_{cm} =  \left( 1 + \frac{  1 }{ f(r)} -  \frac{ L_1 L_2} { r^r f(r) } \right) + 
$$
\be
\sqrt{ \frac{ 1 }{ f(r)} - \left( 1 + \frac{L_1^2}{r^2 f(r)}\right) }  \sqrt{ \frac{  1}{ f(r)} - \left( 1 + \frac{L_2^2}{r^2 f(r)}\right) } 
 \ee
If the above expression is expanded for $ f(r) \ra 0 ( r  \ra 2)$, it simplifies to,
$$
E_{CM} = \frac{1 + \sqrt{ 1 - \frac{L_1^2}{r^2}} \sqrt{ 1 - \frac{L_2^2}{r^2}} - \frac{ L_1 L_2}{r} }{f(r)} +
$$
\be
1 + \frac{ L_1^2 +L_2^2 - 2 r^2 } { 2 r^2  \sqrt{ 1 - \frac{L_1^2}{r^2}} \sqrt{ 1 - \frac{L_2^2}{r^2}}}
\ee
The first term explodes when $f(r) \ra 0$ unless $ L_1^2 = L_2^2 =4$. But, from the effective potentials, the particle will reach the horizon  only when $ L_1^2 < 4$ and $L_2^2 < 4$.  The second term is finite for all values of $L_1^2$ and $L_2^2$ as long as they are small or equal to 4. Hence the possibility exists for CM energy to become very large in this case if the angular momentum of the two particles are fine tuned to be  smaller than $2$.

\subsection{ $e \neq 0$ and $ L =0$ }
In this case, the effective potential becomes,
\be
V_{eff} = \left( 1 - \frac{2}{r}\right) - \left( 1 - \frac{ e Q}{ r} \right)^2
\ee
and $V_{eff}=0$ at 
\be
r = \frac{ ( e Q)^2}{ 2 ( e Q -1)}
\ee
The derivative of the potential, $V_{eff}' =0$ at
\be
r_m = \frac{ ( e Q)^2}{ 2 ( e Q -1)}
\ee
Hence, when $e Q > 1$, there is a maximum for $V_{eff}$ and it is at a positive $r$ as shown in Fig.7. When $e Q <1$, the maximum occur at a negative $r$ value. When $ e Q =1$, $ r_m \ra \infty$. Hence, $ e Q =1$ is the margin of having a root for $V_{eff}$ or not.  When $ e Q  \leq 1$, $V_{eff} < 0$ for all $r > 2$ which facilitates the falling of the particle to the horizon.

\begin{center}
\scalebox{.9}{\includegraphics{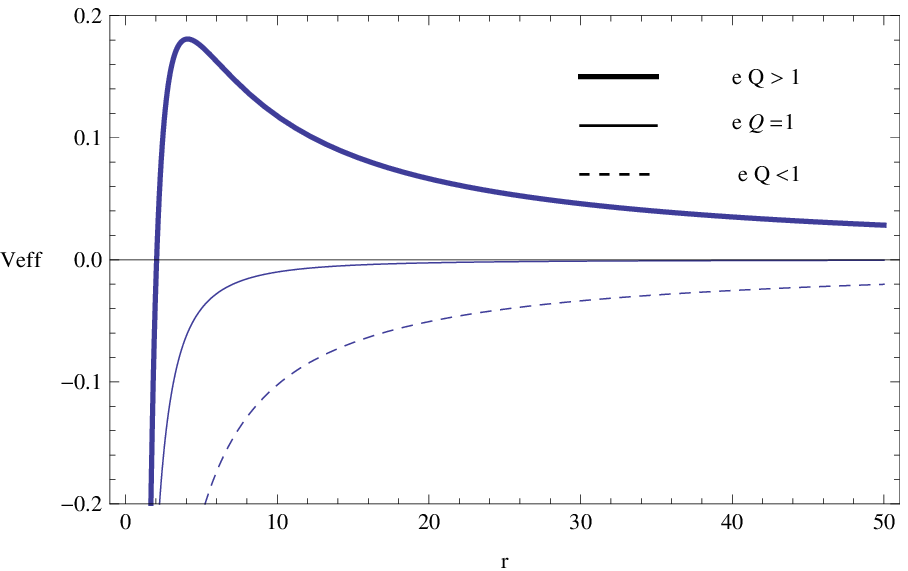}}\\
\vspace{0.1cm}
\end{center}
Figure 7. The graph shows the graph  of $V_{eff}$ vs $r$ for the extreme black hole with $M = 1$ and $ b = 2$ with charged particles without angular momentum.\\

Now, the CM energy for this case, is exactly the same as for the black hole case discussed in section$\refb{cmforlzero}$. The reason is that the angular momentum couples to $R^2(r)$ term in the expression for $E_{CM}$ and since $L=0$ in this case, it is imaterial if it is the extreme black hole or the non-extreme black hole. Hence when the expansion is done around $f(r) \ra 0$, the final result is the same as in section$\refb{cmforlzero}$ as
\be
E_{CM} = 1 + \frac{ 1}{2} \left[ \left( \frac{ 2 - e_1 Q)}{ 2 - e_2 Q} \right) + \left( \frac{ 2 - e_2 Q)}{ 2 - e_1 Q} \right) \right]
\ee
From the above expression, the only way $E_{CM}$ could go to infinity would be either $e_1 Q = 2$ or $e_2 Q =2$. But from the discussion on the effective potential,  $e_1Q  \leq 1$ and $e_2 Q \leq 1$ for either particles to reach the horizon. Hence, $E_{CM}$ will not reach   infinity in this case.

%%%%%%%%%%%%%%%%%%%%

\subsection { $ e \neq 0$ and $ L \neq 0$ }
\be
V_{eff} = \left( 1 - \frac{2}{r}\right) - \left( 1 - \frac{ e Q}{ r} \right)^2 + \frac{ L^2}{r^2}
\ee
When $ r \ra \infty $, $ V_{eff} \ra \infty$ and, $V_{eff}=0$ at 
\be
r_z = \frac{ ( e Q)^2 - L^2}{ 2 ( e Q -1)}
\ee
Hence, the particle will reach the horizon only if $r_z < 2$ which leads to,
\be \label{elratio}
( 2 - e Q)^2 < L^2
\ee
Hence we will study whether the particle will fall into the black hole or not by varying the value of $e Q$.\\

{ \bf Case 1: $eQ < 1$}\\

In this case $ r_z >0$ only if $ ( e Q)^2 > L^2$. In this case  the potential is positive until $ r = r_z$ as given in the thin line in Fig.8. Therefore, the particle cannot fall from far towards the black hole.

When $ L^2 > (e Q)^2 $, $ r_z < 0$. Hence there is no positive root   and the  potential looks like the thick line in Fig.8. Clearly the particle will not fall into the black hole. When $ L^2 = ( e Q)^2$,  $ r_z = 0$ and the potential looks similar to the thin line and the particle will not fall into the black hole.

\begin{center}
\scalebox{.9}{\includegraphics{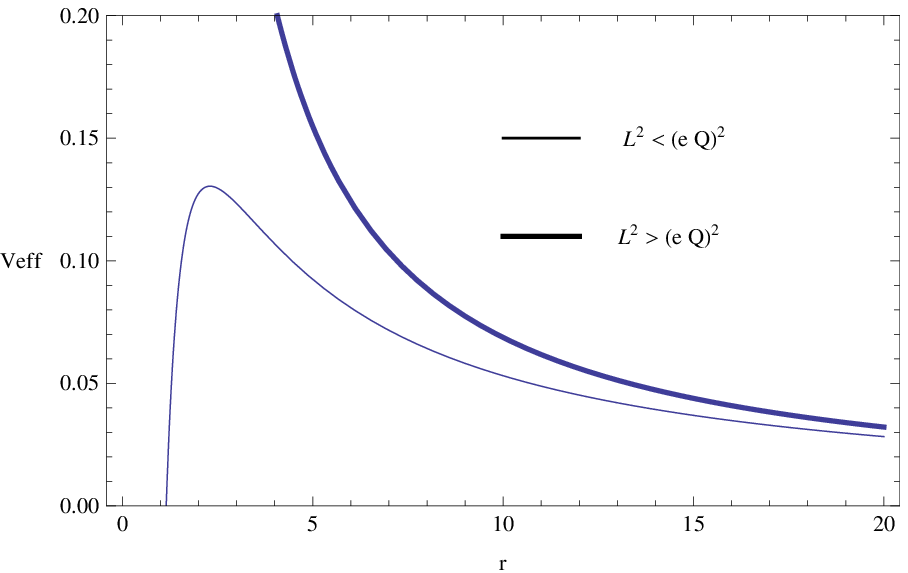}}\\
\vspace{0.1cm}
\end{center}
Figure 8. The graph shows the graph  of $V_{eff}$ vs $r$ for the extreme black hole with $M = 1$ and $ b = 2$ with charged particles with angular momentum. Here $ e Q < 1$.\\

{ \bf Case 2: $e Q =1$}

In this case, 
\be
V_{eff} = \frac{ L^2 -1}{ r^2}
\ee
If $ L \neq 1$, then $ r_z \ra \infty$.  When $ L <1$, the potential becomes negative and when $ L >1$, the potential becomes positive as shown in Fig. 9. When $ L =1$, the potential is zero and there is no gravitational effects on the particle. Hence, one can conclude that when $ L < 1$, the particle can fall into the black hole.

\begin{center}
\scalebox{.9}{\includegraphics{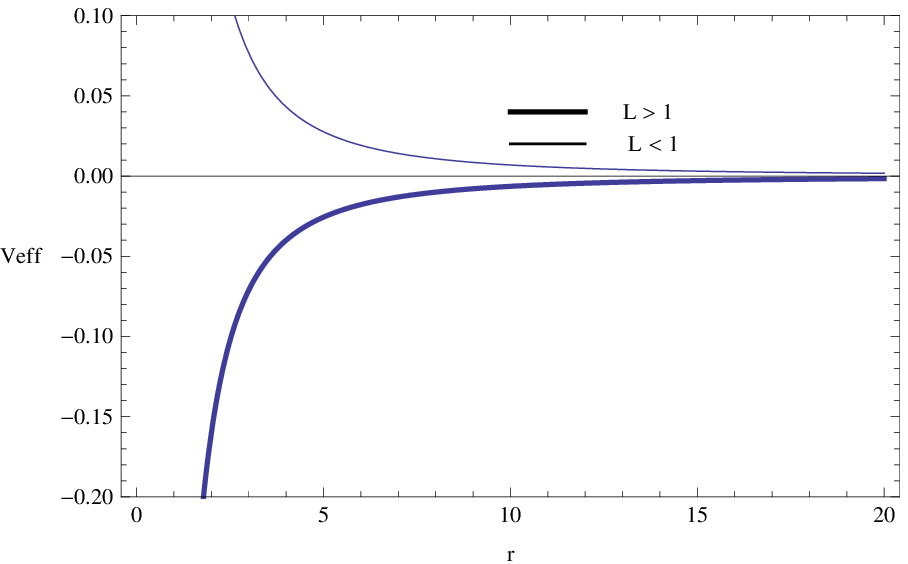}}\\
\vspace{0.1cm}
\end{center}
Figure 9. The graph shows the graph  of $V_{eff}$ vs $r$ for the extreme black hole with $M = 1$ and $ b = 2$ with charged particles with angular momentum. Here $ e Q =1$.\\

{\bf Case 3: $ e Q >1$}

In this case $ r_z >0$ only if $ L^2 > ( e Q)^2$.  How ever $ r_z < 2$ only if  $ L^2 < ( 2 - e Q)^2$. In the thin  graph in the Fig. 10,  
this condition is satisfied and the particle will fall into the black hole. For the small dashed curve,  $ L^2 = ( 2 - e Q)^2$ and $ r_z = 2$. In the thick curev,  $ L^2 > ( 2 - e Q)^2$ and the particle will turn away from the black hole at $ r > 2$ and will not fall into the black hole. For the large dashed curve, $ L < e Q$ which lead to a definite condition for the particle to fall into the black hole.

\begin{center}
\scalebox{.9}{\includegraphics{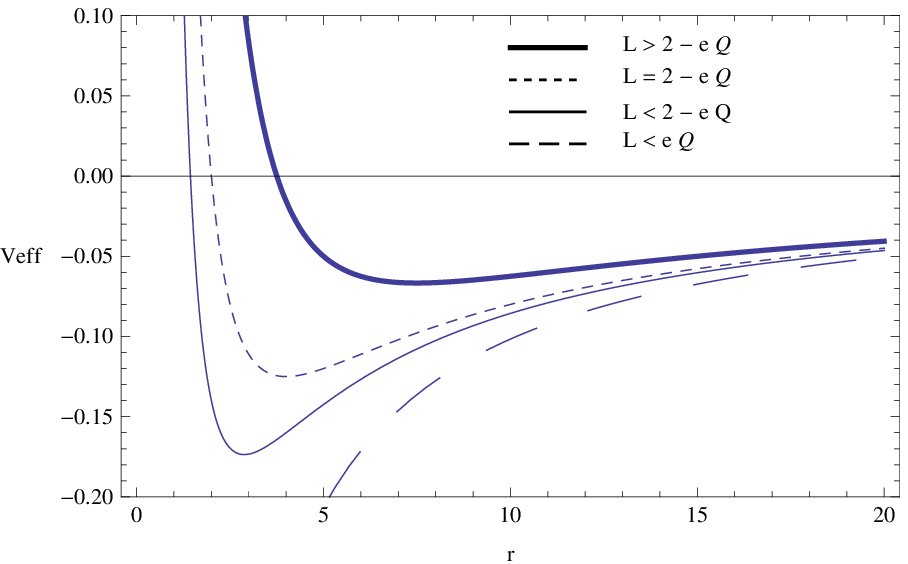}}\\
\vspace{0.1cm}
\end{center}
Figure 10. The graph shows the graph  of $V_{eff}$ vs $r$ for the extreme black hole with $M = 1$ and $ b = 2$ with charged particles with angular momentum. Here $ e Q >1$.\\

The  CM energy in this case is given by,

$$
E_{cm} =  \left( 1 + \frac{ ( 1 - \frac{ e_1 Q}{r}) ( 1 - \frac{ e_2 Q}{r})}{ f(r)} -  \frac{ L_1 L_2} { r^2 f(r) } \right) + 
$$
\be 
\sqrt{ \frac{ ( 1 - \frac{ e_1 Q}{r})^2}{ f(r)} - \left( 1 + \frac{L_1^2}{r^2 f(r)}\right) }  \sqrt{ \frac{ ( 1 - \frac{e_2 Q}{r})^2}{ f(r)} - \left( 1 + \frac{L_2^2}{r^2 f(r)}\right) } 
 \ee
When $E_{CM}$ is expanded around $ f(r) \ra 0 ( r \ra 2)$, the following expressionis obtained.
\be
E_{CM} = \frac{ A}{ f(r)} + B 
\ee
where,
\be
A = \frac{1}{4} \left( ( 2 - e_1 Q) ( 2 - e_2 Q) - L_1 L_2 - \sqrt{ ( 2 - e_1 Q)^2 - L_1}  \sqrt{ ( 2 - e_2 Q)^2 - L_2} \right)
\ee
and,
\be
B = 1 + \frac{\left( ( 2 - e_1 Q )^2 - L_1^2 + ( 2 - e_2 Q)^2 - L_2^2 \right)}{ 2 \sqrt{ ( 2 - e_1 Q)^2 - L_1}  \sqrt{ ( 2 - e_2 Q)^2 - L_2}}
\ee
From the above expressions for $A$ and $B$, , the square roots will be real only if $ ( 2 - e_1 Q)^2  > L_1$ and $   ( 2 - e_2 Q)^2  > L_2$. In that case, $A/f(r)$ will go to infinite leading to large value for $E_{CM}$ for $ f(r) \ra 0$.  From the discussion given above, if $ e Q  \geq 1$, and $ L^2 < ( 2 - e Q)^2$, then the particle will fall into the black hole. Hence large $E_{CM}$ is possible. 

%%%%%%%%%%%%%%%%%%%%%%

%%%%%%%%%%%%%%%%%%%%%%%%
\section{Conclusion}

In this paper we have studied the possibility of large  center-of-mass energy ($ E_{CM}$) of two particles colliding near the horizon of a charged black hole in string theory.  The black hole considered here is  well known as the Gibbons-Maeda-Garfinkle-Horowitz-Strominger (GMGHS) black hole which is static and electrically charged. It has a horizon at $ r = 2M$ and a singularity at $ r = b$.

The $E_{CM}$ and the geodesics of the particle motion were studied in detail. We analyzed these in two cases: first we studied them around the general black hole with $ 2 M > b$. Second, we studied the motion for the extreme case where $ 2 M = b$. The studies were done by changing the electric charge $e$ and the angular momentum $L$ of the particles in motion.

We conclude that for $ 2 M > b$,  $E_{CM}$ will not become infinite. Even if there is a possibility of $E_{CM}$ becoming infinite, the particle will not reach the horizon in such cases. In the extreme case, where $ 2 M = b$, it is possible for the $E_{CM}$ to be infinite in two cases: when $e =0,  L \neq 0$ case and  $ e \neq 0, L \neq 0$ case. In both these cases, if the parameters are chosen appropriately, the $E_{CM}$ would become infinite near the horizon.

As an extension, it would be interesting to study what would  happen if the singularity moves out of the horizon forming a naked singularity with $ b > 2 M$.  Such studies has been done in an interesting paper for the Janis-Newmann-Wincour naked singularity in \cite{patil}.

%%%%%%%%%%%%%%%%%%%%%%%%%

\end{document}